\documentclass[a4paper,fleqn,usenatbib]{pazhbabel}
\usepackage{fullpage}
\usepackage{graphicx}
\usepackage{morefloats}
\usepackage{lscape}
\usepackage{fancyhdr}
\usepackage{amsmath}
\usepackage{caption}
\usepackage[mathscr]{eucal}

\def\Me{\dot{\mathscr{M}}}
\def\Be{\mathscr{B}}
\def\Ce{\mathscr{C}}
\def\Fe{\mathscr{F}}

\begin{document}

\journalinfo{2017}{43}{10}{677}[689]

\title{Modeling of High-Frequency Variability in X-ray Binaries with Black Holes}
\author{\bf A.~N.~Semena
\email{san@iki.rssi.ru}
\address{1}, M.~G.~Revnivtsev\address{1}, T.~I.~Larchenkova \address{2}, A.~A.~Lutovinov \address{1}
\addresstext{1}{Space Research Institute Russian Academy of Sciences, Moscow, Russia}
\addresstext{2}{Astrospace center, P.~N.~Lebedev Physical Institute, Russian Academy of Sciences, Moscow, Russia}} \shortauthor{} \shorttitle{} \submitted{\today} \thispagestyle{empty}

\begin{abstract}
        The properties of the aperiodic variability in X-ray binaries with black holes are considered.
The power spectra of the luminosity variability for a flat accretion disk that is an emission source with a power-law energy spectrum have been modeled. 
At low frequencies the derived power spectrum has the form of a power law with a slope $\rho\approx-1$ and a cutoff at a frequency approximately equal to the fluctuations characteristic frequency at the disk inner edge; at higher frequencies the power spectrum has a complex form. 
The high-frequency variability is suppressed due to the arrival time delays of the photons emerging in the different parts of the disk. 
The presence of azimuthal accretion rate fluctuations in the disk and the azimuthal non-uniformity of the disk surface brightness in the observer's imaginary plane caused by the relativistic effects give rise to an additional variability at frequencies $\sim$200~Hz.

\englishkeywords{
accretion, X-ray binaries, aperiodic variability. 
}

 \end{abstract}

\label{firstpage}

\section{INTRODUCTION}

Strong brightness variability is observed in all systems with disk accretion \cite[see, e.g.][]{scaringi15}. 
This variability is generally believed to be associated with the accretion rate variability in the accretion disk. 
The model of propagating fluctuations \citep{lyubarskii97} was proposed to explain the origin of the power spectrum of the accretion rate variability in the disk. 
In this model the accretion rate variations result from random, independent (at each radius) viscosity parameter fluctuations. 
The propagation of the surface density fluctuations in the $\alpha$-disk can be described by the viscous diffusion equations \citep{lynden74}. 
\cite{lyubarskii97} showed that the accretion rate fluctuations produced in the inner parts of the disk are multiplicatively superimposed on the fluctuations produced in its outer regions.
The model of propagating fluctuations explains well a number of observed phenomena in systems with disk accretion, for example, the lognormal flux distribution \cite[see, e.g.,][]{uttley05, scaringi13} and the time delay of the hard X-ray emission from the inner accretion flow with respect to the soft X-ray emission from the outer regions of an accreting system \citep{pridehorski79, kotov01, revnivtsev10}. 
In particular, this model predicts that accretion rate variability with a power spectrum with the form of a power law $P(\nu)\propto\nu^{\varrho}$ with the slope $\varrho\approx-1$ is generated at the inner parts of an infinite disk \cite[see also][the model of propagating fluctuations in a finite-size disk]{titarchuk07, zdziarski09}.

The shape of the luminosity variability power spectrum observed in many accreting systems, indeed, can be described by a power law with a slope $\varrho\approx-1$, that transforms to a power law with a slope $\varrho<-2$ on the high frequencies.
The frequency at which this transition occurs, so-called break frequency $\nu_b$, is usually associated with the variability near the disk inner edge.
The power spectrum at the frequencies lower than the break frequency is a consequence of the propagating fluctuations in the accretion disk. 
At present, there is no unambiguous answer to the question of what determines the characteristic variability frequency for each disk radius. 
There is evidence both for the assumption that this frequency is the local Keplerian frequency \citep{revnivtsev09} and suggesting that it corresponds to the local viscous time \citep{miyamoto92, icdem11} or even the coronal cooling time \citep{ishibashi12}.
Recent studies \citep{scaringi15} have shown that the break frequency in the luminosity power spectrum for a broad class of systems (active galactic nuclei, low-mass X-ray binaries, cataclysmic
variables, and young stellar objects) is lying on the fundamental plane of system parameters: the accretor mass $M$ , the accretion rate $\dot{M}$ and the inner disk radius $R_{\rm in}$. 
It should be noted that this frequency depends most strongly on the radius of the inner disk edge ($\nu_b \propto R_{\rm in}^2$).
Since the flow structure in different systems can differ significantly, this dependence may suggest that the break in the power spectrum is proportional to the Keplerian frequencies.

In this paper we model the power spectrum of the luminosity variability for an accretion disk rotating around a black hole. 
In our calculations we take into account the relativistic effects that manifest themselves in additional arrival time delays of photons in the disk inner parts and in the nonuniformity of the disk surface brightness in the observer's imaginary plane. 
We also investigate the shape of the derived power spectra as a function of the variability frequency at each disk radius. 
The accretion rate fluctuations are assumed to propagate in the disk retaining their shape and amplitude.

	In the considered model matter is accreted onto the black hole with mass $M=10M_\odot$, specific angular momentum $a=0.6$ and accretion rate $\dot{M} = \frac{1}{4}\dot{M}_{\rm edd} =
    2.5\times10^{-8}M_{\odot}$/yr, the disk rotates in a plane perpendicular to the black hole rotation axis. 
We suppose that the system is observed with the $pi/4$ inclination angle.
The chosen parameters roughly correspond to the parameters of the black hole in the X-ray system Cyg X-1 \citep{gerlinski99, orosz11}.
	
    An extended hot corona is believed to be the source of the  hard X-ray emission in compact binary systems with black holes \cite[see, e.g.,][]{sunyaev79}. 
This hard X-ray emission exhibits strong variability, while the soft X-ray emission associated with the geometrically thin $\alpha$-disk remains weakly variable \citep{churazov01}.

It is worth noting that there exist different views on the possible geometry of the hot corona in which soft X-ray photons are Comptonized to form the hard X-ray spectrum \cite[see discussion in the ][]{nowak02}. 
In particular, some calculations show that the Comptonization of the soft X-ray emission in a geometrically thin corona covering the disk probably cannot give rise to a sufficiently hard X-ray power-law spectrum \cite[see, e.g.,][]{stern95, dove97}. 
At the same time, there is evidence that in active galactic nuclei photons are Comptonized in a geometrically thin layer partially covering the disk flow \citep{wilkins12}. 
In addition, based on the observed frequency-dependent time delays between the energy bands, the model of fluctuation propagation in a geometrically thin flow is most preferred \citep{uttley2014}.

In this paper we will calculate the luminosity variability power spectra for a flat two-dimensional disk, assuming that energy spectum of its emission is described by a power law $I^{p}(\nu)\propto \nu^{-\Gamma}$ with a photon index $\Gamma = 1.5$. 
For the local viscous time, the diffusion time of fluctuations and the local disk brightness we will use the solution for an $\alpha$-disk \citep{ss73} generalized for the Kerr metric \citep{tn74}.

\section{ACCRETION DISK SURFACE BRIGHTNESS AND PHOTONS ARRIVAL TIME DELAYS}

    To construct the model power spectra of the X-ray emission from an accretion disk, it is necessary to calculate the accretion disk surface brightness in the observer's imaginary plane and the photons arrival time delays. 
The emission from the inner regions of a rotating accretion disk will be subjected to the following effects: the Doppler effect, the gravitational redshift, and the gravitational focusing. 
The first two effects determine the observed surface brightness of the disk, while the gravitational focusing determines its observed angular size \citep{cunningham73}. 
If the surface brightness of the accretion disk in its rest frame is constant, then the set of listed effects leads to a nonuniform brightness distribution of the accretion disk in azimuthal
coordinate in the observer's imaginary plane. 
If the disk surface brightness is nonuniform in angular coordinate $\phi$, for example, hot spots are formed on the disk, then flux variations with time will be observed in the observer's rest frame.

The goal of this section is to calculate the momenta of the photons emitted in the vicinity of a black hole at various points of the geodesic and to calculate the photon density in a solid angle
element at the observer's imaginary plane.

\subsection{Photons equation of motion}

As has been noted above, we consider the accretion disk around the rotation black. 
Therefore, the space-time is described with the Kerr metric \citep{kerr}. 
In the subsequent calculations we will use geometric units with $G=c=1$.
To numerically solve the equations, of photon motion written in Boyer-Lindquist coordinates ($t$, $r$, $\theta$, $\phi$) \citep{boyer68} normalized to the black hole mass M, we used the approach proposed by \cite{sharp81} and \cite{zakharov94}. 
It consists in replacing the second-order differential equations describing the motion of photons along the radial and $\theta$-coordinates by four first-order equations. 
To write the equations of motion in a form convenient for numerical calculations, we used the affine parameter described by the following equation: \cite[see, e.g.,][]{zakharov94}:
\begin{equation}
d\sigma = \frac {E d\lambda} {M(r^2 + a^2 \cos^2{\theta})},
\end{equation}
where $d\lambda$ is the affine parameter normalized so that $d/d\lambda = p$ ($p$ is the 1-form of the photon momentum)),
$E = p_{t}$ is the photon energy at infinity.

The system of equations of photon motion can then be written as
\begin{equation}	
\begin{array}{lll}
dr/d\sigma &= & r_{1}\\
dr_{1}/d\sigma &= & 2r^3 + r(a^2 - \xi^2 - \eta) + \\ && ((a - \xi)^2 + \eta)\\
d\theta/d\sigma &= & \theta_{1}\\
d\theta_{1}/d\sigma &= &  -a^2 \cos{\theta}\sin{\theta} + \xi^2 \cos{\theta}/\sin^3{\theta}\\
d\phi/d\sigma &= & -(a - \xi/\sin^2{\theta}) + \\ && a(r^2 + a^2 - \xi a)/ (r^2 - 2Mr + a^2)\\
dt/d\sigma &= & -a(a \sin^2{\theta} - \xi) + \\ && (r^2 + a^2)(r^2 + a^2 - \xi a)/ \\ && (r^2 - 2Mr + a^2),\\
\end{array} \label{eq:motion}
\end{equation}
where  $\xi = L_z/E$ and $\eta = Q/E^2$ are Chandrasekhar's constants, $L_{z}=p_{\phi}$ -- $z$-is the z component of the angular momentum, and $Q$ is Carter's constant \citep{carter68}.
Note that four integrals of motion of a particle ($E$, $L_{z}$, $Q$, $m = p_{i}p^i$) completely define its motion; the photon mass is $m = 0$. 
The system of equations was solved by the eighth-order Runge-Kutta method.

\subsection{Initial Conditions for photons}

To determine the local velocities of the matter in the disk and its surface brightness, we use the generalization of an $\alpha$-disk for the Kerr metric \citep{tn74}. 
The matter in the disk at each radius is assumed to rotate with the local Keplerian angular velocity
\begin{equation}
d\phi/dt = M^{1/2}(r^{3/2} + aM^{1/2})^{-1}
\end{equation}
\cite[see, e.g.,][]{bardeen72}. 
For the subsequent calculations it is necessary to specify the photon emission angles in the disk plane. 
For this purpose, we use an approach that consists of two successive steps. 
In the first step, we calculate approximate photon momenta in the disk plane by calculating the photon trajectories toward the black hole from a distant observer by the so-called backward ray tracing. 
In the second step, accurate momenta of a photon moving toward the observer are found from the derived approximate photon momenta in the disk plane.

To find approximate photon momenta in the disk plane, in the observer's imaginary plane, located at a distance $r_s = 500 r_g$ from the black hole (where $r_{\rm g}$ is the gravitational radius), we specify a spatial grid with $256\times256$ knots. 
This grid covers an area of (150$\times$80)$r_g$ in the observers imaginary plane. 
From the grid knots we calculate the photon trajectories using the affine parameter $d\sigma^{'} = - d\sigma$ , i.e., solve the system of equations of motion (\eqref{eq:motion}).

We assume that the spacetime curvature may be neglected at such a distance (501$r_g$) from the black hole. 
The direction of motion of the photons located at the specified grid knots then coincide with the direction of the line of sight of the distant observer.
The Eulerian coordinates of the grid knots were taken to be 
\begin{eqnarray}
z & = & r_s \cos{\alpha} + b_{y}\sin{\alpha}\nonumber  \\
y & = & b_{x}\nonumber \\
x & = & r_s \sin{\alpha} - b_{y}\cos{\alpha}. \nonumber
\end{eqnarray}
The observer is in the $ZY$ plane, with an inclination angle $\alpha$ relative to the $Z$ axis. 

To simplify the writing, let us introduce two additional parameters, $R_s = \sqrt{x^2 + y^2 + z^2}$ and $r_b = \sqrt{b_x^2 + b_{y}^2}$.
From the momenta in Eulerian space $p^z = \sin{\alpha}$; $p^y = 0$; $p^x=\cos{\alpha}$ we then calculate the momentum components in spherical coordinates:
\begin{eqnarray}
p^\theta & = & (zxp^x - r_b^2 p^z)/(r_b R_s^2)\nonumber,\\
p^\phi & = & -y p^x/r_b^2\nonumber,\\
p^r & = & (xp^x + yp^z)/R_s. \label{eq:eutosp}
\end{eqnarray}

We assume that the spatial momentum components in the Boyer-Lindquist coordinate system are equal to the corresponding components in the spherical coordinate system. 
The last momentum component ($p^0$) is specified from the condition $m=0$:
\begin{equation}
        p^0 = \sqrt{(p^{\phi} g_{t \phi}/g_{t t})^2 - g_{\beta \beta}(p^{\beta})^2/g_{t t}} - p^{\phi} g_{t \phi}/g_{t t},
\end{equation}
where $\beta=(r, \theta, \phi)$, $g_{\gamma \beta}$ are the components of the Kerr metric tensor in Boyer-Lindquist coordinates.

It is important to note that the photons whose momenta are specified in this way move with a deflection from the direction toward the observer by $1/r_s$ at large distances from the black hole. 
Nevertheless, the approximate photon momenta in the disk plane obtained in this way can be used to rapidly find the integrals of motion for the photons moving toward the observer at an infinite distance from the black hole.

Given the momentumcomponents, Chandrasekhar's constants needed to solve the system of equations of photon motion(\ref{eq:motion}) can be determined.
Thereafter, the coordinates of the intersection of the calculated photons geodesic with the disk plane are found. 
Note that the coordinates of the photons in the disk plane whose trajectories were calculated by the method described above are distributed nonuniformly.

In the second step, using the previously found approximate photon momenta in the disk plane, we calculate the photon emission angles ($\theta_p$, $\phi_p$) in the rest frame of the moving disk matter and Chandrasekhar's constants corresponding to the photon motion at infinity toward the observer with a specified accuracy. 
In our calculations we established the level of residuals $10^{-8}$. 

The photon emission angles in the local Lorentz rest frame of the matter are determined at the knots of a spatial $128\times128$ grid on the disk surface. 
This grid covers the disk plane from the innermost stable orbit to 50$r_g$. 
The grid is chosen with a linear step in azimuthal coordinate and with a logarithmic one in radial coordinate. 
At the grid knots we specify a local Lorentz coordinate system, namely a locally nonrotating frame \citep{bardeen72}, which in the disk plane has the following form:

\begin{eqnarray}
dx^0 & = & r\sqrt{\frac{\Delta}{A}} dt \nonumber, \\
dx^1 & = & \frac{r}{\sqrt{\Delta}} dr \nonumber, \\
dx^2 & = & r d\theta \nonumber, \\
dx^3 & = & -\frac{2a}{\sqrt{A}}dt + \frac{\sqrt{A}}{r}d\phi , \label{eq:lnrf}
\end{eqnarray}
where $A=(r^2 + a^2)^2 - a^2\Delta$ and $\Delta = r^2 - 2r + a^2$.
Obviously, the photon momentum components in coordinates $(x^0,x^1,x^2,x^3)$ are expressed via the photon momentum coordinates in Boyer-Lindquist coordinates. 
The momentum components of a photon emitted in the local Lorentz rest frame of the matter are transformed to the momentum components in the coordinates of the local Lorentz frame (\ref{eq:lnfr}) via the Lorentz transformation matrix. 
To construct this matrix in the local Lorentz frame $(x^0,x^1,x^2,x^3)$ we calculate the three-dimensional velocity of the disk matter. 
Thus, from the momentum components in the local Lorentz rest frame of the disk matter we can find the photon momentum components in the Boyer-Lindquist coordinate system and corresponding Chandrasekhar's constants.

For each set of photon emission angles in the local Lorentz rest frame of the matter we can find the corresponding direction of photon motion at infinity ($\theta^o$, $\phi^o$) by calculating the geodesics. 
Below we give expressions for the angles ($\theta^o$, $\phi^o$) written via the momentum components in Eulerian coordinates and the corresponding momentum components in spherical coordinates:

\begin{equation}
\begin{array}{lll}
\cos{\theta^o} &=& p^z/|p|, \\
\sin{\phi^o} &=& p^x/(|p|\sin{\theta^o}), \\
|p| &=& \sqrt{(p^x)^2 +  (p^y)^2 + (p^z)^2},\\
p^z &=& \cos{\theta} p^{r} - r_f \sin{\theta}p^{\theta},\\
p^x &=& \cos{\theta}p^{\theta} (r_f\cos{\phi_k} - \\ &&a\sin{\phi_k}) + \sin{\theta}(\cos{\phi_k}p^{r} -\\&& (r_f\cos{\phi_k} + a\cos{\phi_k})(p^\phi + (a/\Delta) p^r)),\\
p^y &=& \cos{\theta} p^{\theta} (r_f\cos{\phi_k} + \\ &&a\sin{\phi_k}) + \sin{\theta}(\cos{\phi_k}p^{r} +\\&& (r_f\cos{\phi_k} - a\cos{\phi_k})(p^\phi + (a/\Delta) P^r)) .\\ \label{eq:outangles}
\end{array}
\end{equation}
Here, $\phi_k$ is the azimuthal angle of the spherical coordinate system defined by the expression:
\begin{multline}
\phi_k = \phi_f - \frac{a}{2\sqrt{1 - a^2}} \cdot \\
\ln{ \left[ \frac{(r_f - 1 + \sqrt{1 -
a^2})(r_0 - 1 - \sqrt{1-a^2})}{(r_0 - 1 + \sqrt{1 - a^2})(r_f - 1 - \sqrt{1 - a^2})}\right] },
\end{multline}
where $r_f=10^4 r_g$ is the distance from the black hole to the point of the trajectory to which the photon geodesic is calculated, $r_0$ is the radius at which the photon trajectory crosses the disk
plane, i.e., the radius of the corresponding grid knot, $\phi_f$ is the azimuthal photon coordinate in Boyer-Lindquist coordinates at the point of the trajectory at distance $r_{f}$ from the black hole. 
The emission angles $\phi^o$ and $\theta^o$ are calculated with an accuracy of $1/r_f^2$ and $1/r_f^3$, respectively. 
Thus, for $r_f=10^4r_{g}$ the accuracy that we specified for the angles of photon motion at infinity is achieved.

As was already noted above, after the first step of our calculations, we obtained the approximate values of photons Chandrasekhar's constants $\xi$ and $\eta$ in the disk plane and the photon emission angles in the local Lorentz rest frame of the matter (${\theta^p}'$, ${\phi^p}'$)  for a specified velocity field of the matter in the disk. 
The angles specifying the direction of photon motion at infinity (${\theta^o}'$, ${\phi^o}'$), that were calculated using Eqs.\ref{eq:outangles}, and that generally are not equal to the specified direction toward the observer, correspond to these photon emission angles in the local Lorentz rest frame of the matter.

Newton's numerical method is used to determine the angles($\theta^p$, $\phi^p$) corresponding to the direction of motion toward the observer ($\theta^o$, $\phi^o$) with a specified accuracy of $10^{-8}$. 
In this method the residuals ($\theta^o - {\theta^o}'$, $\phi^o - {\phi^o}'$) are reduced through a successive iterative shift in the space of angles ($\theta^p$, $\phi^p$):
\begin{equation}
({\theta^{p}}'';{\phi^p}'') = ({\theta^{p}}';{\phi^{p}}') + (J\cdot (\theta^o - {\theta^o}';\phi^o - {\phi^o}'))^T , \label{eq:newton}
\end{equation}
where ${\theta^p}'$ and ${\phi^p}'$ are the estimates for the photon emission angles in the rest frame of the matter, ${\theta^o}'$ and ${\phi^o}'$ are the emission angles corresponding to these estimates in the observer's rest frame at an infinite distance from the black hole, $J$ is the Jacobian of the mapping $(\theta^p, \phi^p) \rightarrow (\theta^o, \phi^o)$:
\begin{equation}
J = \begin{bmatrix}
\frac{d \theta^p}{d \theta^o} & \frac{d\theta^p}{d \phi^o}\\
\frac{d \phi^p}{d \theta^o} & \frac{d\phi^p}{d \phi^o}
\end{bmatrix}.\label{eq:jac}
\end{equation}

The Jacobian was calculated numerically at each succeeding computational step by shifting  the angles $\theta^p$ and $\phi^p$ in three different directions by $10^{-5}$. 
Finally, by successively reducing the residuals (Eq.\ref{eq:newton}), we obtained the photon emission angles in the local Lorentz rest frame of the matter corresponding to the photon direction toward an infinitely distant observer with a specified accuracy.

\subsection{Calculation of the observed disk surface brightness}

Let us now estimate the radiation energy flux $I^o$ in a solid angle element $\sin(\theta^o) d\theta^o d\phi^o$ from a disk area $\Delta r \Delta \phi$ in the observer's rest frame. 
The observed flux is affected by the gravitational redshift and gravitational focusing. 
The gravitational focusing was determined using the previously calculated Jacobian $J$:
\begin{equation}
d\theta^{p}\wedge d\phi^{p} \approx \det{J}^{-1} d\theta^o \wedge d\phi^o . \label{eq:area}
\end{equation}
To determine the energy flux in the rest frame of the distant observer, we used quantities invariant with respect to the Lorentz transformations. 
The radiation energy flux from the disk toward the observer can be written as 
\begin{equation}
\begin{split}
I^p(\nu^p)f(\theta^p, \phi^p) F(r) dS \cos{\ae} \sin{\theta^p} d\theta^p d\phi^p = \\
I^p(\nu^o)(\nu^p/\nu^o)^3 f(\theta^p, \phi^p) F(r) \sqrt{g_{rr}g_{\phi \phi}} \Delta r \Delta \phi \\
\cos{\ae} ~\det{J}^{-1} (\sin{\theta^p}/\sin{\theta^o}) \sin{\theta^o} d\theta^o d\phi^o , \label{eq:Ip}
\end{split}
\end{equation}
where $dS = \sqrt{g_{rr} g_{\phi\phi}} \Delta r \Delta\phi$ is the emitting area in the Boyer-Lindquist coordinate system, $I^p(\nu^p)$ is the radiation spectrum in the rest frame of the disk matter,
$\nu^o$ and $\nu^p$ are the photon frequencies in the rest frame of the distant observer and the disk matter, respectively, $f(\theta^{p}, \phi^{p})$ is a function defining radiation flux in the $\theta^p$, $\phi^p$ dirrection in the matter rest frame, $F(r)$ is the disk surface brightness in the Boyer-Lindquist coordinate system (in this paper we used the \cite{tn74} solution), $\cos{\ae}$ is the cosine of the angle between the normal to the disk plane and the direction of photon emission measured in the Boyer-Lindquist coordinate system.
Note that $I(\nu)/\nu^3$ and $dS\cos{\ae}$ are invariants with respect to the Lorentz transformation \cite[see, e.g.,][]{rybicki79}.

As a result, from eq.\ref{eq:Ip} we obtain 
\begin{equation}
\begin{array}{lr}
I^o(\nu^o,\theta^o, \phi^o) &= I^p(\nu^o)(\nu^p/\nu^o)^3 f(\theta^p, \phi^p) F(r) \\ & \sqrt{g_{rr}g_{\phi\phi}} \Delta r \Delta \phi  \cos{\ae} \\ & \sin{\theta^p}/\sin{\theta^o}\det{J}^{-1} . \label{eq:flux}
\end{array}
\end{equation}

The cosine of the angle between the direction of photon emission and the normal to the disk can be determined as follows:
\begin{equation}
\cos{\ae} = p^{\theta} \sqrt{g_{\theta \theta}/|p|^2} .
\end{equation}
Here, $|p|^2$ is the length of the spatial part of the momentum squared:

\begin{eqnarray}
|p|^2 &=& -(p^{0})^2 g_{tt} - 2g_{t\phi}p^{\phi}p^0 g_{\phi t} \nonumber \\&=& E^2(\xi g^{\phi t} - g^{tt})(1-g_{t\phi}(g^{\phi\phi}\xi - g^{t\phi})) \nonumber. \\
\end{eqnarray}

Finally, $\cos\ae$ can be written via the integrals of motion as
\begin{equation}
\cos{\ae} = \sqrt{\frac{\eta r^{-2}}{(g^{t\phi}\xi - g^{tt})[1 - g_{t\phi}(g^{\phi\phi}\xi - g^{t\phi})]}} .
\end{equation}

In this paper we do not consider the strongly  lensed photons, i.e., the photons that are deflected through an angle larger than $\pi$, by assuming their contribution both to the total flux from the disk and to the amplitude of the variability being investigated to be negligible. 
The following arguments can be adduced for this approach. 
First, the amplification of such photons is much smaller than that of weakly focused photons \cite[see, e.g.,][]{bozza10}. 
Second, some of the strongly lensed photons cross the disk plane more than once and, consequently, they can be scattered or absorbed by the disk matter, causing their contribution to the system luminosity to be suppressed.

\begin{figure}
\includegraphics[width=\columnwidth]{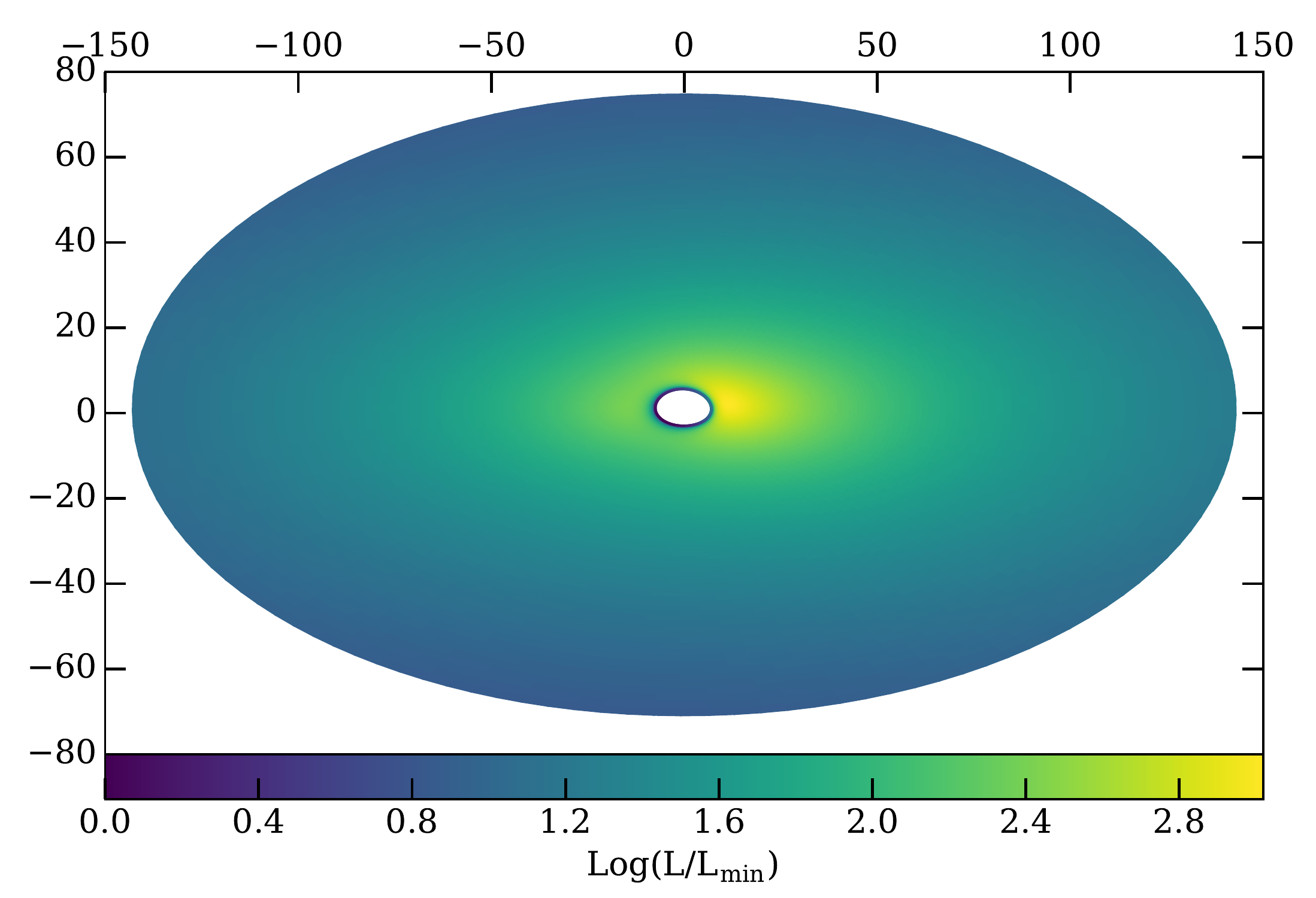}	
\caption{
Map of the luminosity distribution of the geometrically thin disk for the Keplerian velocity field in the distant observer imagenery plane. 
The radiation energy spectrum of each disk elements in the local Lorentz rest frame is assumed to have the form of a power law with a photon index $\Gamma=1.5$. 
The impact parameters in units of $\frac{1}{2}r_{g}$ are along the horizontal and vertical axes.}
\label{fig:disk_surface_brightness}
\end{figure} 	

The luminosity distribution of the parts of the disk at specific wavelength $\nu^0$ can be written as
\begin{equation}
h(\phi, r) = I^o(\nu^o,\theta^o,\phi^o)\sin{\theta^o}, \label{eq:surface_brightness}
\end{equation}
in the distant observer imaginary plane ($I^o$ was calculated for each grid point $r$, $\phi$ on the disk, see Fig.\ref{fig:disk_surface_brightness}).

\subsection{Calculation of the Time Delays}

The time delays of photons at infinity, $\Delta t$, were determined through the numerical calculation. 
For this purpose, we calculate the reference geodesic for which the integrals of motion were specified to be $\xi = 0$ and $\eta = 0$. 
Along this reference geodesic we calculated the time of photon motion $t_{\rm rf}$ to a distance $r_f$ from the black hole. 
Thereafter, for all geodesics calculated in the second step (see above) we determined the convergent integral (follows from the six-th equation of the system (\ref{eq:motion}))
\begin{eqnarray}
\Delta t^0 & = \int_{r_{f}}^{\infty} \frac{-a(a\sin^{2}{\theta^o} - \xi) + (r^2 + a^2)P/\Delta}{\sqrt{P^2 - \Delta((\xi - a)^2 + \eta)}} - \nonumber \\ & \frac{\sqrt{r^4 + a^2r^2 + 2a^2r}}{\Delta} dr .
\end{eqnarray}
This integral corresponds to the time delay at an infinite distance from the black hole between the photon on the reference trajectory and the photon with the integrals of motion $\xi$ and $\eta$. 
The angle $\theta^o$ in the integrand is taken to be equal to the value at infinity. 
To take into account the additional time shifts due to the change in the direction of photon motion as it recedes from the black hole, we additionally introduced corrections of the third infinitesimal order:
\begin{equation}
\begin{array}{ll}
\Delta t_{r^3} = &  6 a^2 \sin{\theta^o}\cos{\theta^o}\cdot\\ & \sqrt{\eta - \cos^2{\theta^o}(\xi^2/\sin^2{\theta^o} - a^2}) r_f^{-3} .
\end{array}
\end{equation}

We obtained the time delays of all photons relative to the reference photon with an accuracy$\propto 1/r_f^{4}$:
\begin{equation}
\Delta t = t_{f} - t_{rf} + \Delta t^o + \Delta t_{r^3},
\end{equation}
where $t_f$ is the calculated time of photon motion from the disk plane to a distance $r_f$.

\begin{figure}
\includegraphics[width=\columnwidth]{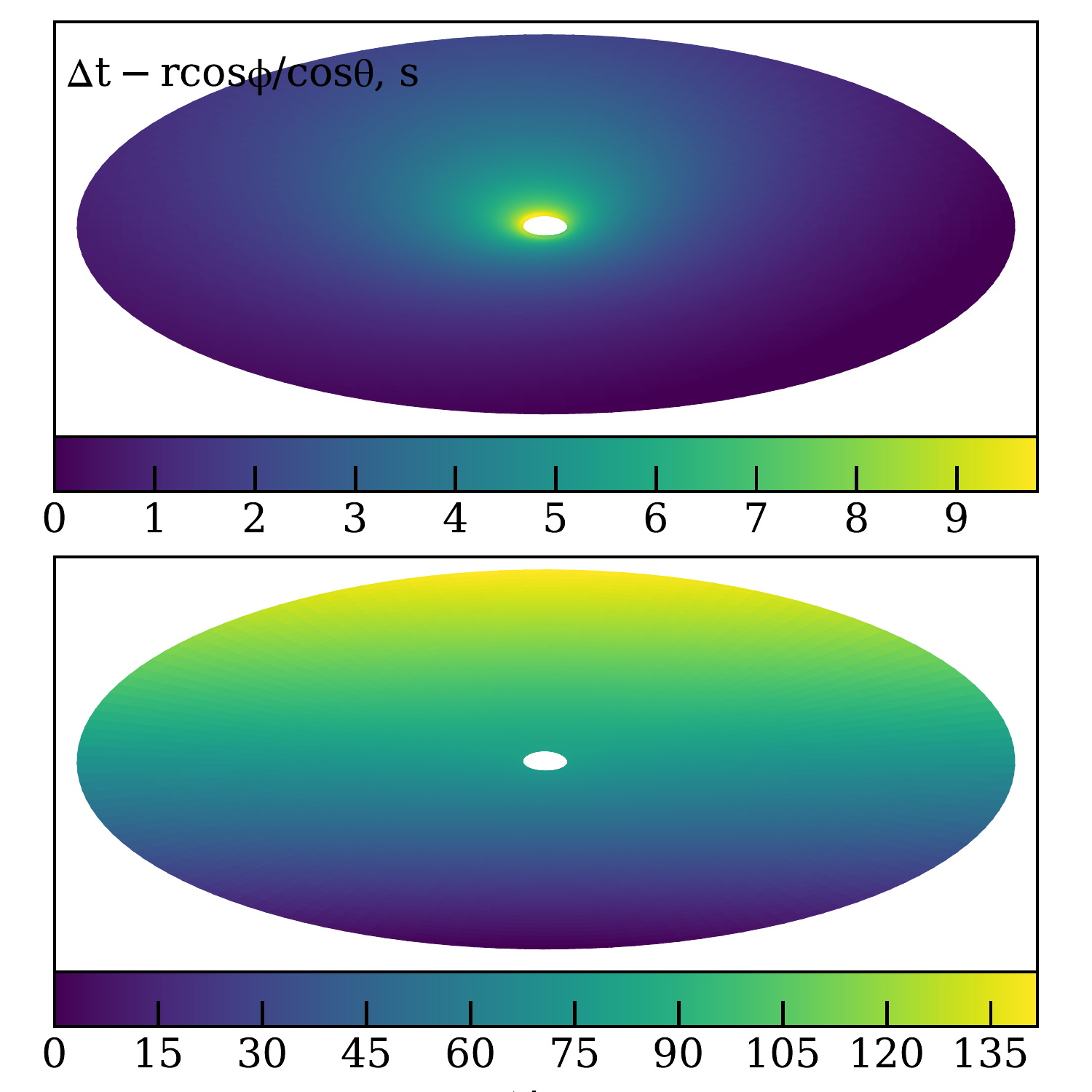}
\caption{
Map of the time delays in the arrival of photons minus the time delays of photon propagation in a flat spacetime (upper panel). 
The original map of the photon time delays (lower panel). 
For both graphs the color diagram is given in units of $GMc^{-3}$. }
\label{fig:disk_time_delays}
\end{figure}

As a result of our calculations, we obtained the distribution of photon arrival time delays along the disk $\Delta t(\phi, r)$ (see Fig.~\ref{fig:disk_time_delays}) that are used together with the observed disk luminosity distribution (Eq. (\ref{eq:Ip})) to calculate the shape of the accretion disk luminosity variability power spectrum.

\section{Formation of the power spectrum}

In the model of propagating fluctuations \citep{lyubarskii97} the accretion rate variability was shown to be generated by a multiplicative superposition of fluctuations on it as the matter propagate to the inner disk radii. 

In accordance with this model, in the finite-difference approach that was used in our calculations the power spectrum of the accretion rate variability for a disk, consisting of a finite number of rings, is formed by a superposition of fluctuations with a certain frequency in ring $i$ on the accretion rate formed in the preceding ring ($i - 1$):
\begin{eqnarray}
\dot{M}(r_i, t) & = & \dot{M}(r_{i-1}, t - \Delta t_{i;i-1}) B(r_i, t)\nonumber \\
& = & \prod_{j=0}^{i} B(r_{j}, t - \Delta t_{ij}) . \label{eq:dotmult}
\end{eqnarray}
In this expression $\Delta t_{ij}$ is the time it takes for the accretion rate fluctuations from radius $r_j$ to reach radius $r_i$ (where $j<i$, and $r_{j}$ > $r_i$), $\dot{M}(r_{j}, t)$ is the accretion rate in ring $j$, $B(r_{j}, t)$ is the variability generated in ring $j$. 
Assuming that the velocity of the matter in the flow does not depend on the accretion rate, i.e., the disk surface density changes little with accretion rate and, on average, remains constant, the propagation of the fluctuations is completely advective: the fluctuation shape and amplitude are retained when moving along the disk.

\begin{figure}
\includegraphics[width=\columnwidth]{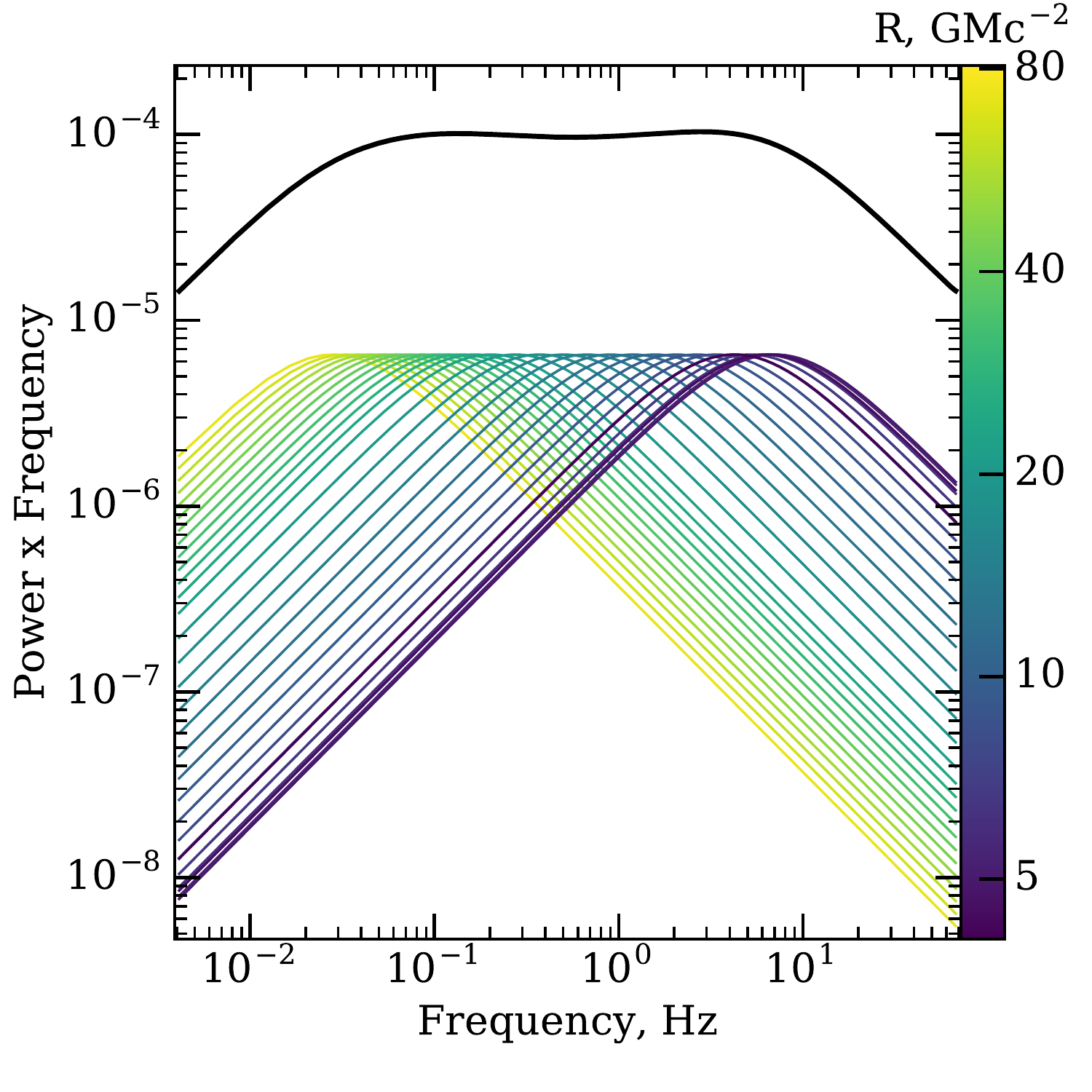}
\caption{
    The solid black line indicates the power spectrum of the accretion rate formed in 30 disk rings with a logarithmic step. 
The color lines indicate the power spectra of the fluctuations in each ring. 
The line color corresponds to the ring radius at which the variability emerged. 
The power spectra of the fluctuations have the form of a wide Lorentzian with zero mean with an amplitude $\sigma$ and a width corresponding to the inverse diffusion time on a given disk ring $-v_{r}/r$. 
The fluctuation amplitude was specified to be the same for all disk rings.}
\label{fig:dotMps}
\end{figure}

The approach described above was used by \cite{arevalo06} to investigate the power spectrum of the variability of the disk luminosity, which was estimated as a sum of the luminosities of all disk rings. 
The luminosity is proportional to the accretion rate, which was determined for each ring in accordance with Eq.~(\ref{eq:dotmult}). 
A random fluctuation $B(t)$ in each ring was generated by the method described in \cite{timmer95}. 
\cite{arevalo06} showed that when choosing the rings with a logarithmic increase in width $\Delta r/r = const$ and the same variability power in each step, the power spectrum of the disk luminosity variability actually has the form of a power law with a slope $\rho\approx-1$.

Subsequently, \cite{ingram13} concluded that the power spectra obtained through a multiplicative superposition of fluctuations could be estimated without generating numerically random accretion rate fluctuations. 
Since the accretion rate in each succeeding ring is formed multiplicatively and is determined from Eq.~(\ref{eq:dotmult}), its Fourier transform is a convolution of the Fourier transform of the fluctuations in ring $i$ and the Fourier transform of the accretion rate in ring $(i - 1)$:
\begin{equation}
\Me(\nu, r_i) = \Me(\nu, r_{i-1}) \otimes \Be(\nu, r_{i}),
\end{equation}
where $\otimes$ is the convolution operation and $\Fe(\nu)$ is the Fourier transform of the function $F(t)$ ($\Fe(\nu) = \int_{-\infty}^{+\infty}F(t)e^{-i 2\pi\nu t}dt$).
The power spectrum of the accretion rate in ring $i$ is a convolution of the power spectra of the accretion rate in the preceding ring $(i-1)$, with the power spectrum of the fluctuations produced in the current ring $i$, because the phases of $\Be(\nu, r_i)$ and $\Me(\nu, r_{i-1})$ are random and independent \citep{ingram13}:
\begin{equation}
|\Me(\nu, r_i)|^2 = |\Me(\nu, r_{i-1})|^2 \otimes |\Be(\nu, r_{i})|^2.
\end{equation}

Thus, the power spectrum of the accretion rate for a disk consisting of $N$ rings can be described with the following expression:
\begin{equation}
|\Me(\nu, r_i)|^2 = \coprod_i^N |\Be(\nu, r_{i})|^2 , \label{eq:nudotm}
\end{equation}
where $\coprod$ is the sequential convolution operation. 
An example of the power spectrum of the accretion rate calculated in this way is shown in Fig.~\ref{fig:dotMps}.

The luminosity variability power spectrum of the ``one-dimensional'' disk with zero photons arrival time delays can be written as
\begin{equation}
\begin{array}{lll}
P(\nu) &=& \sum_{i,j=1}^N h_i h_j \Me(r_i, \nu)^{*}\Me(r_{j}, \nu)\\ &=&
\sum_{j=1}^{N}h_{j}^2|\Me(r_j, \nu)|^2 + \\&&2 \sum_{i=1}^{j-1} h_{i} h_{j} \cos{(2\pi\Delta t_{ij}\nu)}|\Me(r_{i}, \nu)|^2,
\end{array}
\end{equation}
where $\Delta t_{ij}$ are the time delays of the signal rings $i$ and  $j$;  $h_i$, $h_{j}$ are their luminosities, and $^*$ is the complex conjugation operation.

For the ``one-dimensional'' accretion disk with the variability generated on viscous time scales  $t_{\rm visc} = -r/v_{r}$ (in our work $v_{r}$ is the radial velocity of the moving matter) \cite{ingram13} showed that the final power spectrum of the disk luminosity variability has the form of a power law with a slope $\varrho\approx-1$, that is changed to the $\rho\approx-2$ at the frequency corresponding to the inverse viscous time for the disk inner edge. 
In these calculations the power spectrum of the variability for each disk radius had the form of a Lorentzian with zero mean and a width equal to the inverse viscous time:
\begin{equation}
|\Be(\nu)|^2 = \frac{\sigma^2 \nu_{w}}{\pi T(\nu_w^2 + \nu^2)} , \label{eq:lorentznoise}
\end{equation}
where $\sigma$ is the fluctuation amplitude, $T$ is the duration of the light curve for which the power spectrum is calculated, and $\nu_w = 1/t_{\rm visc}$ is the inverse viscous time at disk radius $r$. 
Note that \cite{titarchuk07} obtained an analytical solution of the diffusion equations for the fluctuations in the disk. 
They showed that the power spectrum of the random fluctuations that were generated at some radius and reached the inner disk edge must have the shape of a plateau with a transition on the inverse viscous time scale of this radius to a power law with a slope of $-3/2$ (in the case of a constant kinetic viscosity along the disk). 
In this paper we use a variability with the power spectrum in the form of a Lorentzian due to the necessity of comparing our results with those of other authors \citep{arevalo06, ingram13}. 
Basically, this profile is fairly close in shape to the profile in \cite{titarchuk07}.

In our paper, in contrast to \cite{ingram13}, we calculate the power spectrum of the luminosity variability not for a one-dimensional disk but for a two-dimensional flat disk in the Kerr metric by taking into account the relativistic effects. 
The latter affect the observed spatial disk size and cause the disk surface brightness nonuniformity and relativistic time delays of the emitted photons. 
Furthermore, in addition to the accretion rate variations generated during the motion of the matter in the disk, in our calculations we take into account the nonuniform distribution of the accretion rate in azimuthal coordinate and investigate the shape of the power spectrum as a function of the characteristic variability frequency.
The power spectra of the accretion rate variability within each ring are calculated by the method described above.
	
To calculate the power spectrum of the disk luminosity variability, we specify a two-dimensional $r$, $\phi$ grid with a logarithmic step in $r$-coordinate and a constant amplitude of fluctuations at each grid point. 
In azimuthal coordinate the grid is specified to be uniformly filled $\Delta \phi = {\rm const}$. 
Such a choice of the grid is dictated by the computer simulations of accretion flows, where angular momentum is transferred through the magnetorotational instability \citep{beckwith08}.
In these calculations an equal accretion rate variability amplitude is contained in each disk ring with a logarithmic step in radius. 
The power spectrum of the fluctuations on each ring has the form of a wide Lorentzian with zero mean and a width corresponding to the inverse viscous time (Eq.~(\ref{eq:lorentznoise})).

\begin{figure}
\includegraphics[width=\columnwidth]{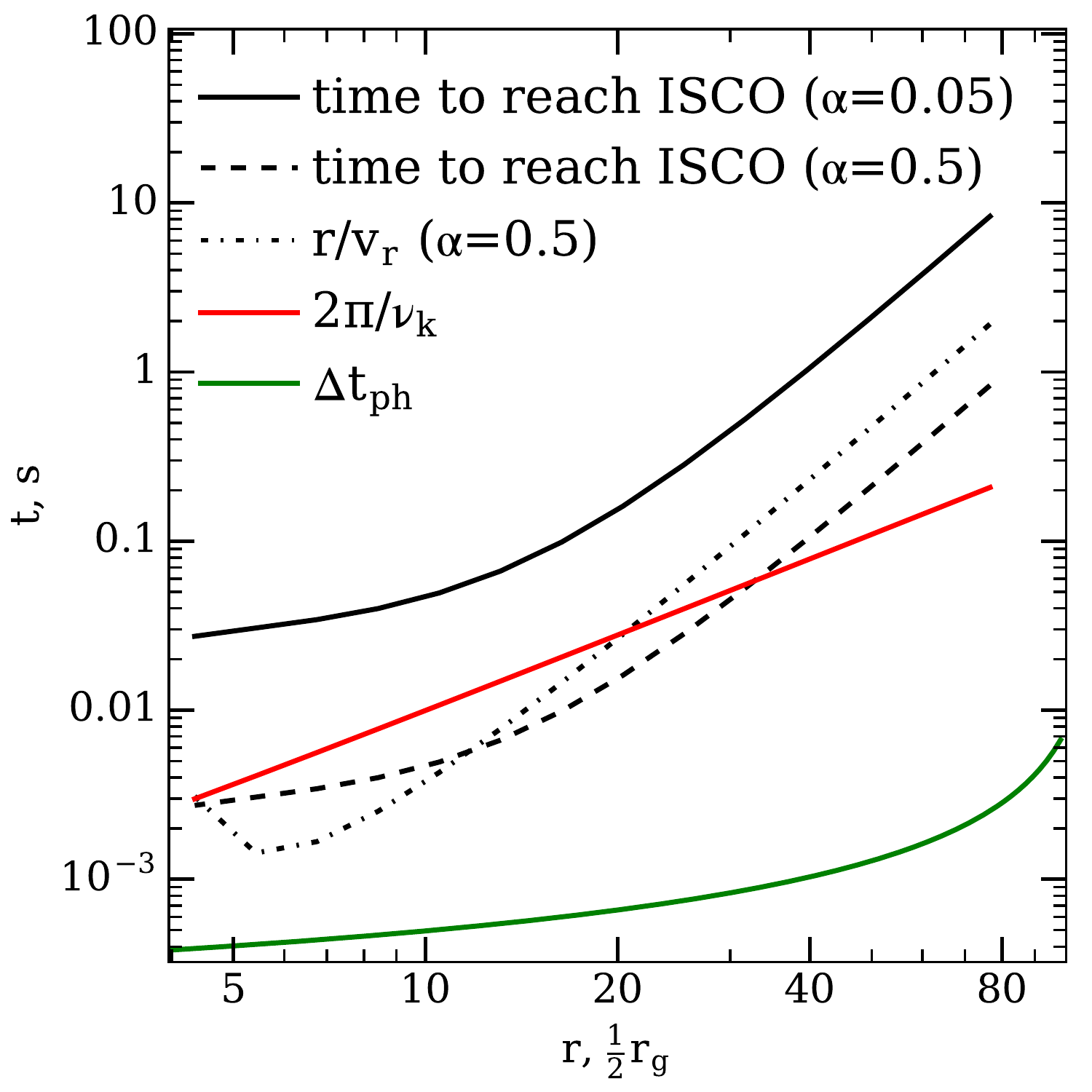}
\caption{Diffusion times ($-r/v_{r}$), the time of motion to the innermost stable orbit for two values of the viscosity parameter$\alpha_{\rm visc}$, the Keplerian time, and the maximum time delay between the arrivals of photons from different parts of the disk from one radius.}
\label{fig:times}
\end{figure}

\begin{figure}
\includegraphics[width=\columnwidth]{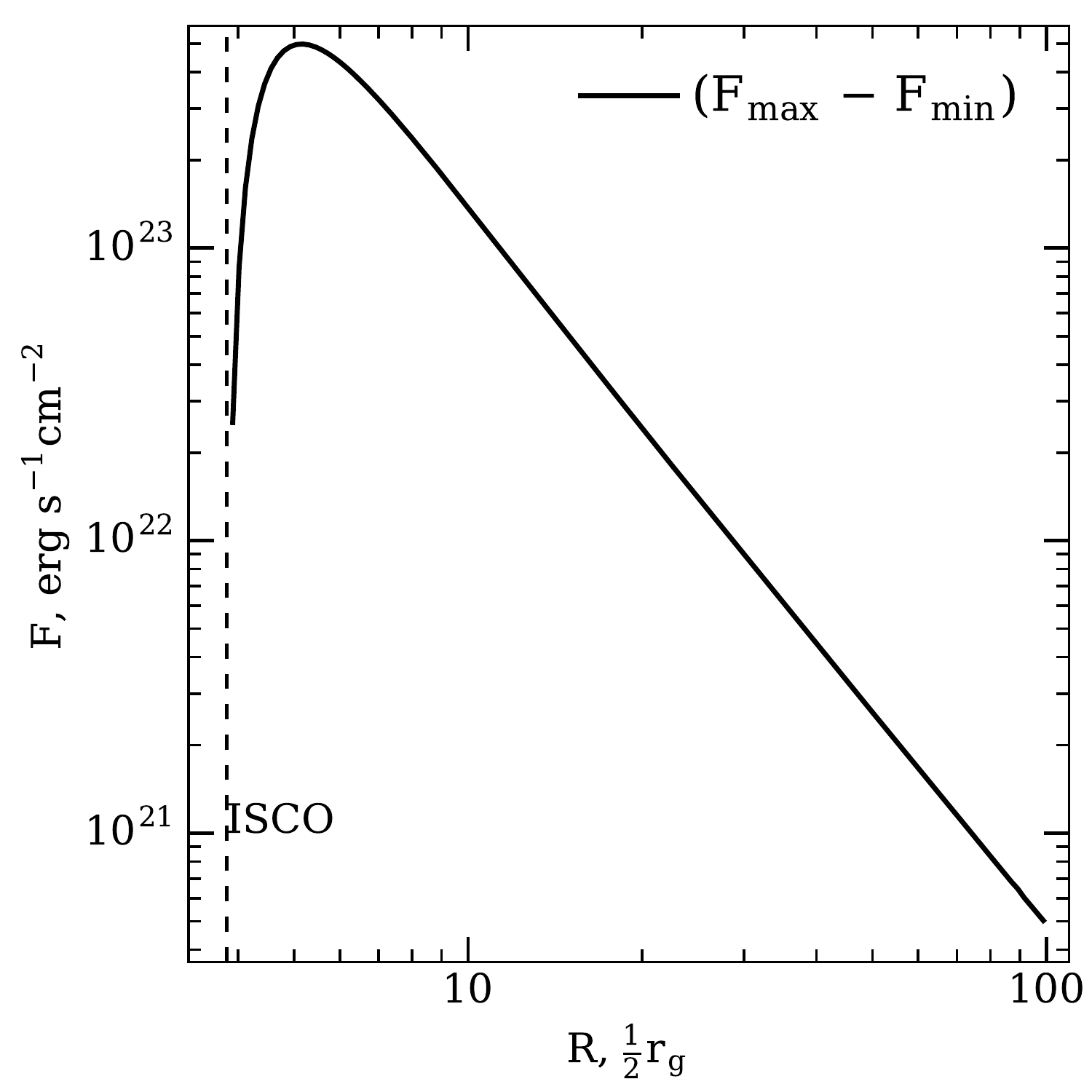}
\caption{Difference between the maximum and minimum disk surface brightnesses for each radius from the viewpoint of a distant observer for the case of a surface brightness corresponding to the \citep{tn74} solution.}
\label{fig:lumty}
\end{figure}

The viscous time $t_{\rm visc} = -r/v_{r}$ and the diffusion time of the fluctuations $\int_{r}^{ISCO}v_{r}^{-1}dr$ are estimated from the Thorne-Novikov solution for a geometrically thin, optically thick, radiation-dominated $\alpha$-disk \citep{ss73}. 
Various characteristic time scales in an optically thick disk are presented in Fig.~\ref{fig:times}. 
The following values of the viscosity parameter were used to find the characteristic time scales of the disk: $\alpha_{\rm visc}=0.05$ and $\alpha_{\rm visc}=0.5$. 
At the chosen mass and accretion rate the disk transition to the radiation-dominated regime occurs at a radius $r\approx 65 r_{g}$.
The radius of the innermost stable orbit ($r_{\rm ISCO}$) at which the disk surface brightness in this solution is zero is $1.92 r_{g}$.

The power spectrum of the luminosity variability for a two-dimensional disk is calculated as follows:
\begin{equation}
P(\nu) = \sum_{l,n}^N h_l h_n \Me_\phi(r_l, \phi_l,  \nu)^{*}\Me_\phi(r_{n},\phi_n, \nu) ,
\end{equation}
where $l$ and $n$ correspond to any two disk elements $(r_l, \phi_l) \in ([R_{\rm out}, 
R_{\rm in}], [0, 2\pi])$, $R_{\rm out}$ is the outer disk radius,
$\Me_\phi(r_n,\phi_n, \nu)$ is the Fourier transform of the accretion rate in ring $n$ calculated from the formula
\begin{eqnarray}
\dot{M}_\phi(r, \phi, t) = \dot{M}(r, t)\cdot C(r, \phi, t),\\
\Me_\phi(\nu, \phi, r) = \Me(\nu, r) \otimes \Ce(\nu, r, \phi),
\end{eqnarray}
where $\dot{M}(r, t)$, $\Me(\nu, r)$ were determined from Eqs.~(\ref{eq:nudotm}) and (\ref{eq:dotmult}), $\Ce(\nu, \phi, r)$ is the Fourier transform of the fluctuation superimposed on the accretion rate in a ring with radius $r$ due to the flow inhomogeneity in coordinate $\phi$. 
This variability is specified to correspond to the accretion rate nonuniformities moving with Keplerian velocities (``spots'' on the disk surface). 
In this case, the variability $\Ce$ has the following properties:
\begin{equation}
\Ce(r_{l},\phi, \nu)^{*} \Ce(r_n, \phi, \nu) = \delta_l^n \label{bprop:1},
\end{equation}
where $\delta_l^n$ is the Kronecker delta;
\begin{equation}
\sum_{i,j=1}^{N_\phi}\Ce(\phi_i,r,\nu)^{*}\Ce(\phi_j,r,\nu) = \delta(\nu) \label{bprop:2},
\end{equation}
where $\delta(\nu)$ is the delta function, $N_{\phi}$ is the number of ring elements;
\begin{equation}
\Ce(\phi_i,r, \nu)^{*}\Ce(\phi_j,r,\nu) = |\Ce(r, \nu)|^2 \cos{\frac{(\phi_i - \phi_j)\nu}{\nu_k}} , \label{bprop:3}
\end{equation}
where $\nu_k$ is the Keplerian rotation frequency at the corresponding radius. 
It follows from the properties (\ref{bprop:1}) and (\ref{bprop:3}) that the fluctuations $C$ has a zero variability amplitude at frequencies below the Keplerian one $|\Ce(0<\nu<\nu_k)|^2 = 0$. Thus, the fluctuations $\Ce$ do not change the mean accretion rate in the ring.

Since the disk surface brightness is nonuniform in angle $\phi$, due to the relativistic effects, a spot rotating in a circular orbit will give rise to a periodic signal (see, e.g., \cite{schnittman04}). 
A large number of spots at different radii will give rise to a continuous power spectrum of the luminosity variability. 
The amplitude of the periodic signal drops with decreasing surface brightness nonuniformity, which is defined as the difference between the maximum and minimum surface brightnesses at each radius.
In the considered case (the disk inclination of $\alpha=\pi/4$ relative to the observer, disk velocity and luminosity are defined with the Thorne-Novikov solution) the dependence of the disk surface brightness nonuniformity on radius is presented in Fig.~\ref{fig:lumty}.

According to the property (\ref{bprop:1}), the cross-correlation of the $\Ce$ components between different rings is disregarded, because the spots rotating on the disk surface are uncorrelated between themselves, i.e., the phase shift between two spots on two adjacent rings changes randomly after a random time interval.

The subsequent calculation of the power spectrum consists of two parts:\\
1. The calculation of the cross-correlation components between rings:
\begin{equation}
P(\nu) = \sum_{l,n}^N h_l h_n \Me(r_l, \phi_l,  \nu)^{*}\Me(r_{n},\phi_n, \nu), \label{eq:pcross}
\end{equation}
where the indices $l$ and $n$ in this case correspond to the elements that do not belong to one ring. 
Since the fluctuations $\Ce$ between any two rings are uncorrelated, they may be disregarded when calculating these components.\\
2. The calculation of the Fourier transform of the light curve for each ring:
$$
\begin{array}{lr}
\Me_{\rm ring}(\nu_q) = & \sum_{j=0}^{N_\phi} \sum_{p=-N_f/2}^{N_f/2} \left( \right.   h_j \Me_p \Ce_{q-p} e^{-i\frac{\nu_p\phi_j}{\nu_k}} \\ & e^{i 2 \pi \nu_q \Delta t(\phi_j) + i\phi_j\nu_q/\nu_k} \left. \vphantom{\sum_{0}^{0}}\right) .
\end{array}
$$
Here, $i$ in the exponent is the imaginary unit, $q$ and $p$ are the indices of the frequency channels in the discrete spectrum, $N_{f}$ is the number of channels in this spectrum. 
The variability power in a ring $|\Me_{\rm ring}|^2 = (\Me_{\rm ring})^{*}(\Me_{\rm ring})$ is a sum of complex {\bf nonrandom} phase. 
Using the property of independence and randomness of the phases of $\Me(\nu)$ at different frequencies, more specifically, $\left[\Me(u)\right]^{*}\Me(v) = \delta(u-v)|\Me(u)|^2$, where $u$ and $v$  are the frequencies, we can write the expression to estimate the power spectrum of the ring luminosity as
\begin{equation}
\begin{array}{ll}
P(\nu) =& \int_{0}^{2\pi}\int_{0}^{2\pi}\int_{-\infty}^{+\infty} |\Me(u)|^2|\Ce(\nu - u)|^2 \\ &  e^{-i\frac{u\Delta\phi_{12}}{\nu_k}}du \\& h(\phi_1)h(\phi_2) e^{i 2 \pi \nu (\Delta t_{12} - \Delta\phi_{12}/(2\pi\nu_k))} d\phi_1 d\phi_2 ,
\end{array}
\end{equation}
where $\Delta\phi_{12} = \phi_1 - \phi_2$, and $\Delta t_{12} = t(\phi_1) - t(\phi_2)$ are the time delays between the arrivals of photons from the ring elements at $\phi_1$ and $\phi_2$.
For a finite number of elements in a ring the power spectrum in the $q$-th frequency channel is
\begin{equation}
\begin{array}{ll}
P_q = & \sum_{ij=1}^{N_{\phi}}h_{i}h_{j} \left( \sum_{p=-N/2}^{N/2} |\Me_p|^2 |\Ce|^2_{q-p}\right. \\ & \cos{(2\pi\Delta t_{ij}\nu_q + \Delta \phi_{ij} (\nu_q - \nu_p)/\nu_k)} \left. \vphantom{\sum_{1}^{2}}\right) \label{eq:pfin},
\end{array}
\end{equation}
where $|\Me_p|^2$ is the amplitude of the discrete power spectrum at frequency $\nu_p$.

In general form the Fourier transform of the fluctuations $C(\phi)$ at a grid point ($r$,$\phi$) is a series of harmonics located at frequencies that are multiples of the Keplerian frequency at a given radius $r$ \cite[see, e.g., the power spectrum for a circular spot from][]{schnittman04}. 
The relations between the harmonic amplitudes and the phase shifts between the harmonics depend on the spot shape. 

As an example, let us choose the simplest case for the shape of a rotating spot where the accretion rate nonuniformity in coordinate $\phi$ has the form of a harmonic function:
\begin{equation}
C(r, \phi, t) = 1 + \epsilon \cos(\phi + 2\pi \nu_k t)
\end{equation}
where $\epsilon$ is the signal amplitude. In this case, the Fourier transform is
\begin{equation}
\Ce(r, \nu) = \delta(\nu) + \epsilon \delta(\nu - \nu_k) e^{-i\frac{\phi \nu}{\nu_k}},
\end{equation}
while its power spectrum is
\begin{equation}
|\Ce(r, \nu)|^2 = \delta(\nu) + \epsilon^2\delta(\nu-\nu_k) ,
\end{equation}

According to (\ref{eq:pfin}), the final power spectrum of the fluctuations in each ring is a sum of the fluctuation power spectrum and the same power spectrum but shifted by the Keplerian frequency:
\begin{equation}
\begin{array}{ll}
P_q = & \sum_{ij=1}^{N_{\phi}}h_{i}h_{j} \left[ |\Me_q|^2 \cos{(2\pi\Delta t_{ij}\nu_q)} + \right. \\ & |\Me_{q-k}|^2 \cos{(2\pi\Delta t_{ij}\nu_q + \Delta \phi_{ij})} \left. \vphantom{\sum_{1}^{2}}\right]) . \label{eq:pdip}
\end{array}
\end{equation}
This simple case corresponds to the largest contribution of spots to the disk variability at low frequencies; therefore, we used the spots with such a shape in our calculations.

The final power spectrum is the sum of the values obtained from (\ref{eq:pfin}) and (\ref{eq:pcross}).

\section{RESULTS}

We calculated the power spectra of the light curves for an accretion disk in the vicinity of a rotating black hole for several possible characteristic frequencies of fluctuations: the frequencies equal to the inverse viscous time for two values of the parameter $\alpha_{\rm visc}$ and the Keplerian frequency.

\begin{figure}

\includegraphics[width=\columnwidth]{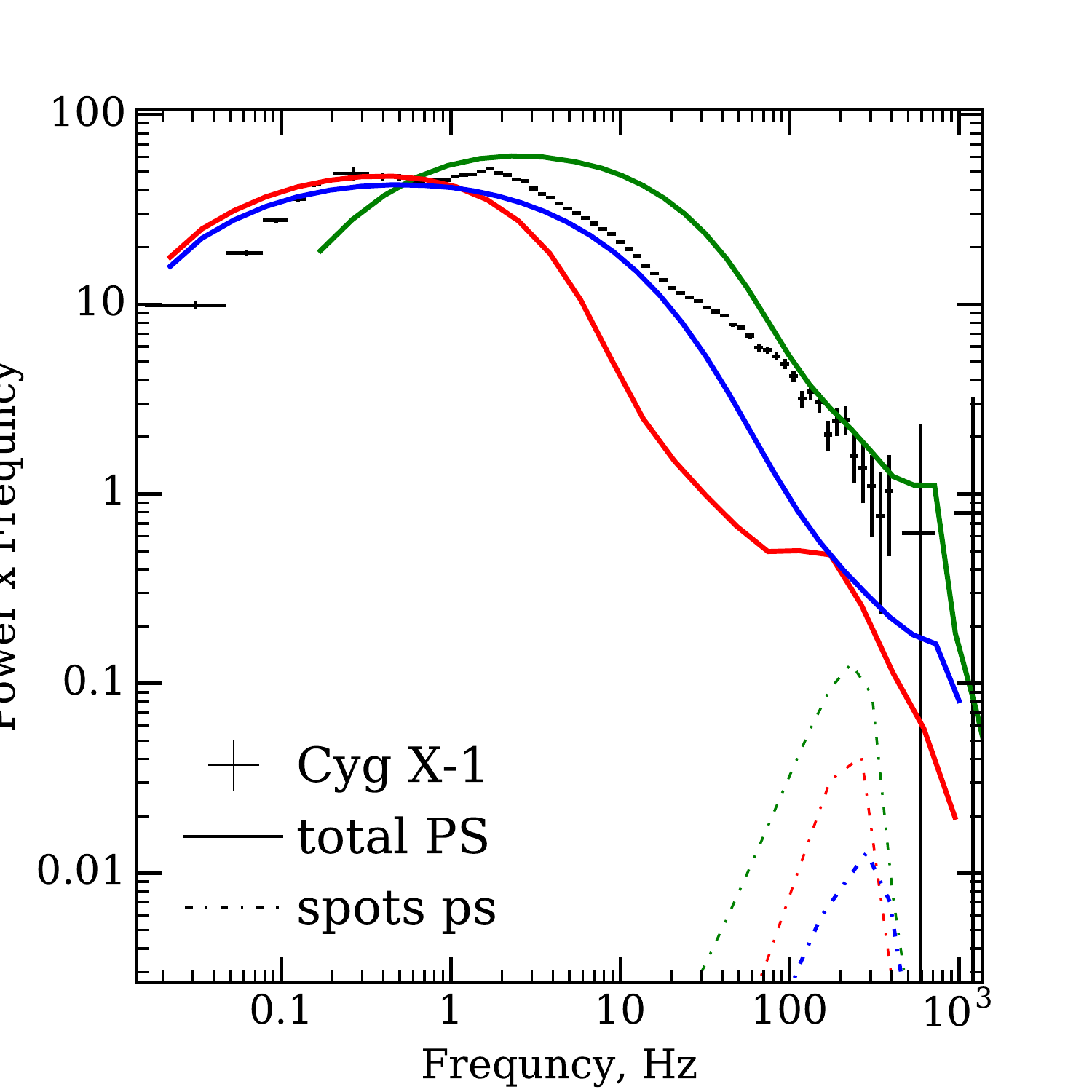}
\caption{
Power spectra of the brightness variability for an accretion disk rotating around a black hole with a mass of $10 M_\odot$ and specific angular momentum $a$=0.6.
The red and green solid curves represent the calculated power spectra for disks with $\alpha_{\rm visc}$, equal to 0.5 and 0.05, respectively. 
The dash-dotted curves of the corresponding colors indicate the variability due to the motion of spots on the accretion disk surface. 
The blue curve corresponds to the power spectrum for a classical$\alpha$-disk with an inner radius of 1.92 $r_g$, calculated without any relativistic effects of strong gravity but with the Doppler shift and the Lorentz aberration of light. 
The black crosses indicate the power spectrum for Cygnus X-1 measured from RXTE data in its hard spectral state in the 2--60~keV energy band.
}
\label{fig:psfin}
\end{figure}

Figure~\ref{fig:psfin} presents the power spectra of the luminosity variability calculated for a two-dimensional flat disk with viscosity parameters $\alpha_{\rm visc}=0.05$ (red curves) and $\alpha_{\rm visc}=0.5$ (green curves). 
The blue curves represent the power spectra for a classical $\alpha$-disk calculated by taking into account the Doppler shift and the relativistic aberration of light in Minkowskiy geometry. 
The solid lines represent the full power spectra of the luminosity variability for a flat disk. 
The dash-dotted lines indicate the contribution of the accretion rate azimuthal fluctuations to the luminosity power spectrum.

For comparison, the figure also presents the power spectrum of the luminosity variability for the X-ray binary Cygnus X-1 in its hard state, obtained from RXTE data. 
The light curves for the power spectrum were extracted in the entire PCA energy band (2--60~keV).
The choice of the energy band is dictated by the recently detected additional instrumental noise in the soft PCA energy subbands \citep{revnivtsev15}.
This noise, which is presumably associated with the change in the efficiency of alternate photon detection in adjacent energy channels on short time scales $\sim 200$~$\mu$s can give rise to some peculiarities in the observed power spectra if they are extracted from the light curves measured in the narrow PCA energy subbands.  

The derived power spectrum can be approximately described by a set of several power laws. 
Each succeeding power law has a steeper slope with increasing frequency. 
The transition from the low-frequency power law with a slope $\varrho=0$ to the power law with $\varrho \approx-1$ occurs at the frequency corresponding to the characteristic variability frequency at the disk outer edge. 
The next transition to a power law with a slope $\varrho\approx-2$ occurs at the frequency lower than the characteristic frequency of fluctuations at the inner disk edge.

The lack of the variability power due to the correlations of the fluctuations in the outer and inner parts of the disk is observed in the frequency range 3--100~Hz for the viscous parameter $\alpha = 0.05$(red curve) and 30--500~Hz for $\alpha$=0.5(green curve). 
This deficit of power gives rise to a structure resembling the peak of a quasi-periodic oscillations at frequencies 100--400~Hz.

It should be noted that the characteristic frequencies increase with the viscosity parameter $\alpha_{\rm visc}$, while the shape of the power spectrum of the luminosity variability changes. 
Thus, the power spectrum of the luminosity variability for a disk with a large viscosity parameter $\alpha_{\rm visc}$ cannot be obtained by simply shifting the power spectrum obtained for a small $\alpha_{\rm visc}$ toward higher frequencies.

   The disk surface brightness nonuniformity caused by the relativistic effects gives rise to the luminosity variability at Keplerian frequencies. 
   This effect, for the inclination of the disk plane $\alpha=\pi/4$, turns out to be most significant at $r=2.72r_{g}$, corresponding to the Keplerian time $\sim4$~ms for $M=10M_\odot$ (see fig \ref{fig:lumty}). 
The component associated with the motion of the spots on the disk surface, which is indicated by the dash-dotted line, is clearly seen in the power spectrum of the disk brightness variability. 
It has a maximum at the frequency $\approx250$~Hz. 
The more spots are located at one disk radius, the greater is the shift of this component toward high frequencies. 
In the case of the classical $\alpha$-disk in Minkowskiy geometry the accretion rate nonuniformity in azimuthal component also manifests itself as an additional variability (the blue dotted curve in Fig. \ref{fig:psfin}). 
This component is associated with the disk surface brightness nonuniformity caused by the relativistic aberration of light and the Doppler shift. 
Its amplitude is predictably smaller than the amplitude of the analogous component for a relativistic disk in Kerr geometry, due to the additional nonuniformity caused by the gravitational lensing.
    
   In our calculation the amplitude of the accretion rate variability in the spots, $\Ce$, was chosen to be equal to the amplitude of the accretion rate variability attributable to stochastic fluctuations $\Be$. 
   It can be seen from Fig.~\ref{fig:psfin} that for $\alpha_{\rm visc}$ > 0.05 the spots variability amplitude in the power spectrum of the disk luminosity is everywhere smaller than the amplitude of the variability related to the accretion rate variations. 
   In the case of a flow with a large viscosity parameter $\alpha_{\rm visc}$ the luminosity variability caused by the disk accretion rate nonuniformity in azimuthal coordinate will not contribute significantly to the power spectrum of the luminosity variability. 
   A flow with a large viscosity is presumably the source of the accretion rate variability in low-mass binary systems \citep{churazov01}. 
   It should be noted that the amplitude of the variability from the azimuthal fluctuations increases with the accretion disk inclination, in contrast to the variability related to the radial accretion rate fluctuations. 

   At frequencies above $\sim$300~Hz the luminosity variability of the flat disk is suppressed. 
   This variability    suppression occurs due to the time delays in the propagation    of photons from different parts of the disk.
   The suppression occurs above the certain frequency and is more pronounced for a flow with a larger $\alpha_{\rm visc}$, i.e., with the shorter fluctuations characteristic time scales.
   As has been noted in the Introduction, the hard X-ray emission most likely results from the Comptonization of seed photons in a hot corona \citep{eardley75, sunyaev79, pozdnyakov83}. 
   The Comptonization process suggests numerous photon scatterings. 
   As a result, such a photon spends additional time to escape the corona \cite[see, e.g.,][and subsequent papers]{kazanas97}. 
   \cite{titarchuk07} showed that at the sufficient accretion rate a wind whose optical depth could reaches 1 should form above the disk. 
   Such wind, along with the corona,  leads to the additional delays in the photons arrival times. 
   In the luminosity power spectrum the variability suppression  through the photon random walk manifests itself as a transition to a steeper power law at the frequencies above the inverse random walk time. 
   The delays in the arrival times of photons emerging in the different disk parts also manifest themselves in the power spectrum in a similar way. 
   It should be noted that the models suggesting extended coronas with long photon diffusion times are inconsistent with the observed luminosity power spectra in different energy bands \citep{lin00}. 
   I also follows from Fig.~\ref{fig:psfin} that the suppression of the variability in the power spectrum of Cygnus X-1, which could be associated with the photon diffusion time in the corona (a break in the power spectrum), is absent at the frequencies below $\sim100$~Hz. 
   This means that the suppression of the variability through the delays in the arrival of photons, which, according to our calculations, is expected at the frequencies above 300~Hz, and the suppression through random photon walks during Comptonization (unobservable at frequencies below 100 Hz) can give comparable effects in the power spectrum at high frequencies.

\section{CONCLUSIONS}
In this paper we obtained the model power spectra of the luminosity variability for an accretion disk extending from 50~$r_g$ to the innermost stable orbit and rotating around a black hole with the mass $10M_\odot$ and specific angular momentum a = 0.6. 
The power spectra have been obtained for various characteristic variability frequencies in the disk. 
The variability amplitude of the geometrically thin disk at high frequencies (above $\sim300$~Hz) was shown to be suppressed due to the photons arrival time delays.

In addition, the power spectrum was found to have the shape of a power law with a cutoff at the variability frequency corresponding to the radius of the disk with its maximum surface brightness in the observer's imaginary plane. 
For the black hole considered in our model this radius is $\sim 3 r_g$. 
The variability above this frequency turns out to be partially suppressed due to the correlation of the accretion rate fluctuations in different parts of the disk. 

Our calculations have shown that an additional variability component in the power spectrum arises from the surface brightness nonuniformity of the disk in azimuthal coordinate in the rest frame of the matter (the spots moving over the disk) and its surface brightness nonuniformity in the observer's imaginary plane caused by the focusing of disk emission near the black hole and the Doppler effect for the matter rapidly rotating in inner orbits. 
The amplitude of this variability in the power spectrum increases rapidly with the frequency up to several hundred Hz and makes no contribution at frequencies below $\sim 30$~Hz. 
The variability amplitude is at a maximum at frequencies $\sim 250$~Hz.

The power spectra were obtained for various values of the characteristic frequency at which variability can be generated at each disk radius. 
We showed that if the propagation time of fluctuations in an $\alpha$-disk is longer than the characteristic times of the fluctuations themselves, then the disk luminosity variability amplitude is retained.
Otherwise (the accretion rate variations have higher frequencies) the brightness variability is suppressed compared to the accretion rate variability. 
In the considered case, where the viscous and diffusion times are determined from the solution for a geometrically thin disk, the variability amplitude is retained at frequencies below the frequency $\nu \sim -v_r/r$ corresponding to a $10 r_g$ scale and is partially suppressed above this frequency.

The power spectrum in this case has the characteristic form of a power law with a slope $\rho\approx-1$ and a cutoff frequency slightly below the inverse viscous time at the disk inner edge. 
The shape of the power spectrum above the break frequency can be described by a power law with a slope $\rho \approx -2$ with a number of peculiarities. 
At frequencies above 300~Hz the disk brightness variability is suppressed due to the arrival delays of the photons emerging in the different parts of the disk.

\bigskip

\section{ACKNOWLEDGMENTS}

This work was supported by the Russian Science Foundation (grant no. 14-12-01287). 
We thank E. Churazov and A. Veledina for the discussion of the paper and their remarks.

\label{lastpage}

\end{document}